# Effective Infection Opportunity Population (EIOP) Hypothesis in Applying SIR Infection Theory

Hiroshi Isshiki, Institute of Mathematical Analysis (Osaka)

Masao Namiki, Former Vice President of Toshiba Corporation

Takeshi Kinoshita, Professor of Emeritus of The University of Tokyo

Ryosuke Yano, Tokio Marine & Nichido Risk Consulting Co. Ltd., Senior Principal Investigator

Abstract:

The SIR infection theory initiated by Kermack-Mckendrick in 1927 discusses the infection in an isolated population with uniform properties such as the uniform population distribution. In the infection, there exist two aspects: (1) The quantitative aspect and (2) the temporal aspect. Since the SIR theory is a mean-field theory, it can't match both aspects simultaneously. If the quantitative aspect is matched, the temporal aspect can't be matched, versa. The infection starts from a cluster, and it spreads to different places increasing the size of the infection. In general, even in the case of the infection in a big city, the infection grows within a limited population. Namiki found and named this kind of population as an effective population. He proposes that if the hypothesis is adopted, the quantitative and temporal aspects can be matched simultaneously.

## 1. Introduction

Since SIR theory initiated by Kermack-Mckendrick in 1927 [1, 2] is a mean-field theory, it discusses the infection phenomena in an isolated population space with uniform distribution. Hence, it can't reflect the effects due to uneven population density.

In the infection, there exist two aspects: (1) The quantitative aspect and (2) the temporal aspect. it can't match both aspects simultaneously. since the SIR theory is a mean-field theory. If the quantitative aspect is matched, the temporal aspect can't be matched, versa.

The infection starts from a cluster, and it spreads to different places increasing the size of the infection.

The flying fire has a certain tendency because it spreads through daily activities such as commuting to school, shopping, entertainment, and meetings. In general, even in the case of the infection in a big city such as Tokyo, the infection grows within an extremely limited population.

This kind of limited population is named "The Effective Infectable Population (EIP)" by Namiki or one of the authors of the present authors. It is proposed by Namiki that the quantitative and temporal aspects of the infection could be matched simultaneously.

## 2. SIR infection theory initiated by Kermack-Mckendrick and the limit [1, 2]

Let $t$ be time, and $N$, $S$, $I$, $R$ be the number of the whole population, uninfected persons (or susceptible persons strictly speaking), infected persons, and recovered persons (including dead). The SIR theory initiated by Kermack-Mckendrick in 1927 is given by

$$N = S + I + R, \qquad (1)$$

$$\frac{dS}{dt} = -\frac{\beta}{N}SI, \qquad (2)$$

$$\frac{dI}{dt} = \frac{\beta}{N}SI - \gamma I, \qquad (3)$$

$$\frac{dR}{dt} = \gamma I, \qquad (4)$$

where $\beta/N$ is the infection ratio. If $\beta/N$ is defined like this, the infection rate is not affected by the size of the population. $\beta$ is called the infection force, and $\gamma$ is the recovery rate.

From Eqs. (1) through (4), the following equations are obtained:

$$\frac{d(S+I+R)}{dt} = 0, \qquad (5)$$

$$\frac{d(I+R)}{dt} = \frac{\beta}{N}SI. \qquad (6)$$

Since Eq. (3) can be rewritten as

$$\frac{dI}{dt} = \gamma\left(\frac{\beta}{\gamma N}S - 1\right)I. \qquad (7)$$

the fundamental reproduction number $R_0$ and the infection threshold $\rho_0(t)$ are given by

$$R_0 = \beta, \qquad (8)$$

$$\rho_0(t) = \frac{\beta S(t)}{\gamma N}. \qquad (9)$$

The fundamental reproduction number is the number of the infected persons infected from one infected at a time.

When one infected appears at $t = 0$, the infection extends or shrinks according to $\beta/\gamma$ bigger or smaller than 1, respectively.

Although the SIR theory is a superior theory capturing the essence, the SIR theory has two problems. One comes from the legal aspect, and the other from the fact that it is a mean-field theory.

Firstly, the legal aspect: the new type of corona-virus was specified as the first kind of designated infectious disease. Since those who tested as positive are isolated even if undeveloped, they do not contribute to the infection. Hence, if Eqs. (1) through (4) are used, the infection does not occur. However, since there are infected persons in the city who are not tested, this causes the infection. Odagaki's SIQR theory that is explained in the next section introduces this effect.

The other problem comes from the fact that the SIR theory is a mean-field theory. This theory is based on that the uninfected and infected persons are equally distributed in the space. Hence, for example, it can't distinguish the case whether the infected persons are distributed uniformly or concentrated in the space. Briefly, it can't reflect the characteristics of the space. Since the real infection phenomena take place in a non-uniform world usually, the non-negligible discrepancy could occur between the SIR theory and real phenomena if the utmost efforts are made.

The nonlinear simultaneous ordinary differential equations given by Eqs. (1) through (4) are integrated numerically using Euler's method. AS shown in Fig. 2, the sufficient accuracy is obtained with $dt$=1day.

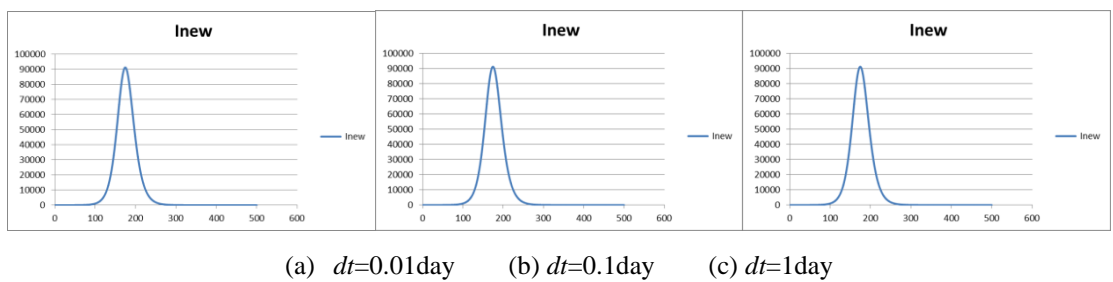

(a) $dt$=0.01day    (b) $dt$=0.1day    (c) $dt$=1day

Fig. 1. Accuracy of Euler's method (Newly infected people $I_{new}$).

## 3. Odagiri's SIQR theory [3]

In Odagaki's SIQR theory, the infected persons $I$ in SIR theory is divided into two groups. One is the infected but not isolated people I in the city and the other is positive by test and isolated people $Q$.

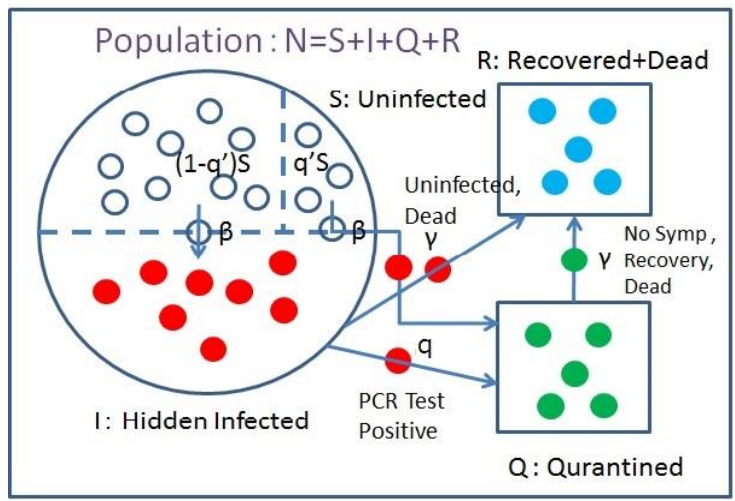

Fig. 2. Illustration of Odagaki'sISQR Model.

The SIQR theory is given by

$$N = S + I + Q + R \qquad (10)$$

$$\frac{dS}{dt} = -\frac{\beta}{N} SI \tag{11}$$

$$\frac{dI}{dt} = (1-q')\frac{\beta}{N} SI - qI - \gamma I \tag{12}$$

$$\frac{dQ}{dt} = q'\frac{\beta}{N} SI + qI - \gamma' Q \tag{13}$$

$$\frac{dR}{dt} = \gamma I + \gamma' Q . \tag{14}$$

The parameters $q$ and $q'$ is defined in Fig. 2. Among the infected people in the city, $qI$ persons are tested in the city and judged as positive. (1-$q'$)$S$ persons remain in the city and involved in the infection in the city, and $q'$ persons check the infection at the hospital.

Since Eq. (12) is rewritten as

$$\frac{dI}{dt} = (p+\gamma)\left(\frac{(1-q')\beta S}{(p+\gamma)N} - 1\right)I , \tag{15}$$

the infection threshold (or effective reproduction number) becomes

$$\rho_0(t) = \frac{(1-q')\beta S(t)}{(q+\gamma)N} . \tag{16}$$

If Eqs. (12) through (14) are summed, the following equation is obtained

$$\frac{dI+Q+R}{dt} = \frac{\beta}{N} SI . \tag{17}$$

The fundamental reproduction becomes $\beta$.

## 4. Comparison with real data and Namiki's "Effective Infection Opportunity Population Hypothesis (EIOP)"

In Fig. 3, the infection situation in Tokyo is shown. From this figure, the two aspects of the infection are obtained:

(1) Qualitative aspect

The peak value of new certified infected and the accumulated infected persons are 200 persons/day and 5,000 persons, respectively.

(2) Temporal aspect

The period from the beginning (middle of March) to the end (middle of May) of the first wave of infection is 60 days.

Since this figure includes artificial factors such as capturing and isolating infected persons by cluster tracing, it may not be possible to perfectly match with SIQR equation, but the basic time transition could be represented by the SIQR equation. However, both the quantitative and temporal aspects cannot be matched simultaneously without assuming a population that is significantly smaller than the actual population of Tokyo, based on a hypothesis such as the effective infection opportunity population. A part of the reason for this will be discussed in the next section.

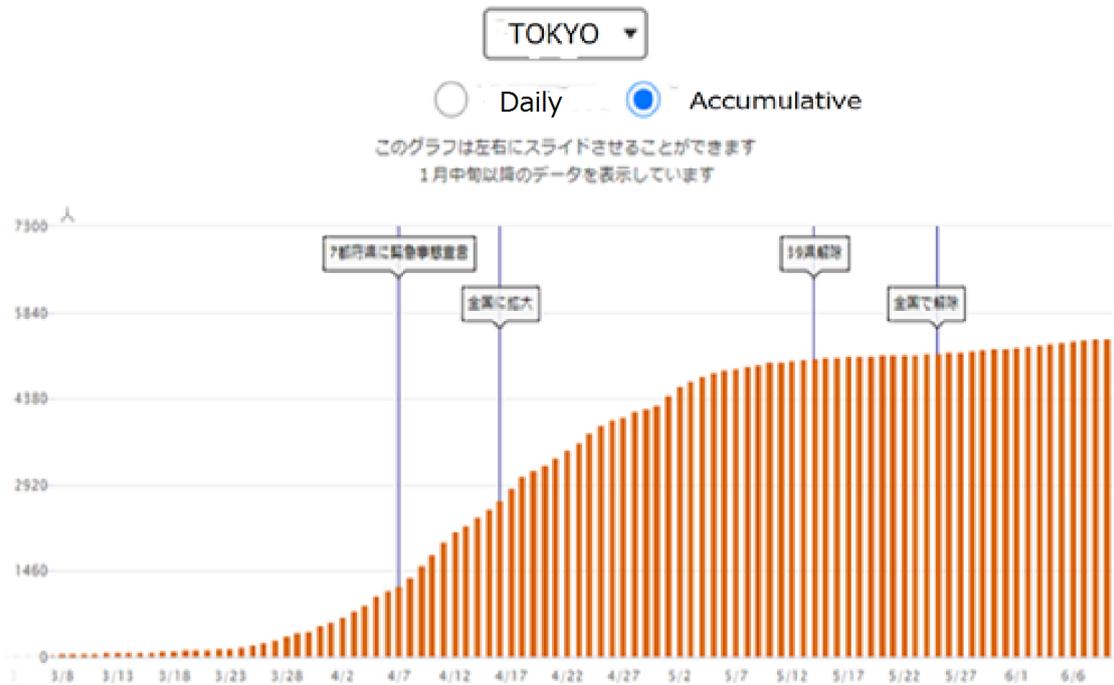

(a) Changes in the cumulative number of infected people

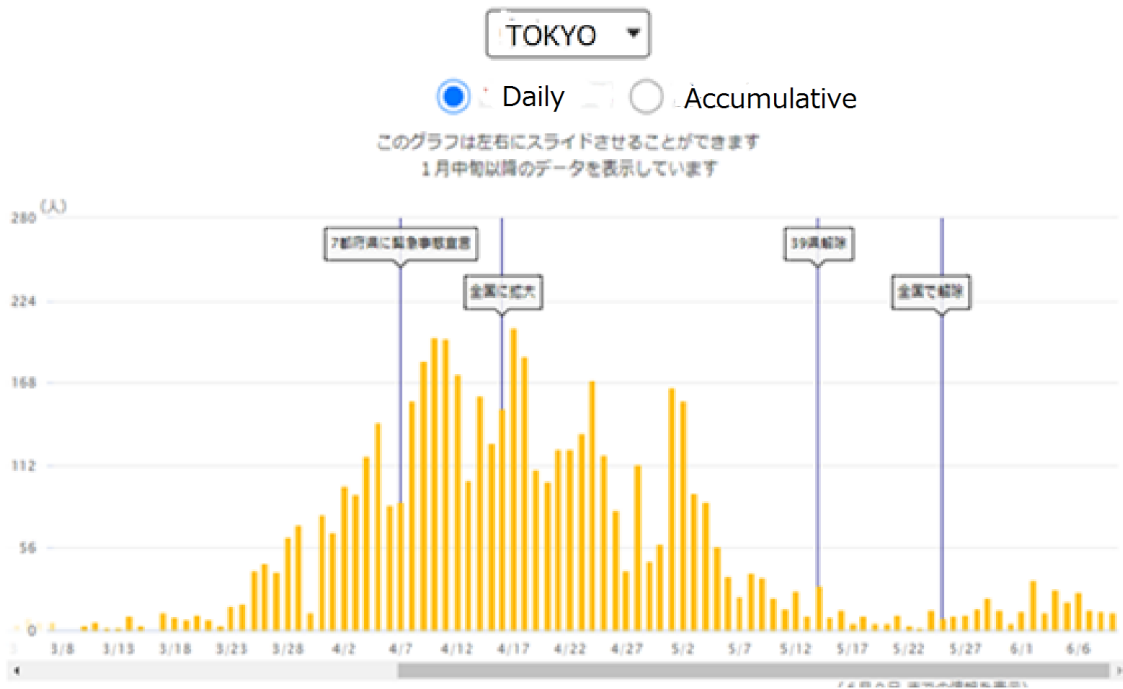

(b) Changes in the number of new infections

Fig. 3. Status of new coronavirus infection in Tokyo

(https://www3.nhk.or.jp/news/special/coronavirus/data/).

If Namiki's hypothesis is introduced into Odagaki's theory, the theory is modified as follows:

$$N = S + I + Q + R \tag{18}$$

$$\frac{dS}{dt} = -\frac{\beta}{N_a} SI \tag{19}$$

$$\frac{dI}{dt} = (1-q')\frac{\beta}{N_a} SI - qI - \gamma I \tag{20}$$

$$\frac{dQ}{dt} = q'\frac{\beta}{N_a} SI + qI - \gamma' Q \tag{21}$$

$$\frac{dR}{dt} = \gamma I + \gamma' Q. \tag{22}$$

In Eqs. (19) through (21), $N$ is replaced with $N_a$ ($< N$). From a different point of view, it is equivalent to multiplying the infection rate $\beta$ by $N/N_a$.

This hypothesis is numerically verified by conducting a series of numerical calculations. The effective infection opportunity population is changed to Na=13,000,000, 50,000, 25,000, 10,000. The infection-force β is also gradually changed.

First, the temporal aspect is adjusted as Na=13,000,000. The parameters used in the calculation are shown below:

Na=13,000,000 ; S[0]=12,999,999; I[0]=1; β=0.8671; γ=0.08 ; q=0.46; q'=0.1

Checking at the calculation results, the infection period is almost the same, but the peak value of the newly infected person $I_{new}$ and the cumulative number of infected persons $I_{acm}$ are not matched at all.

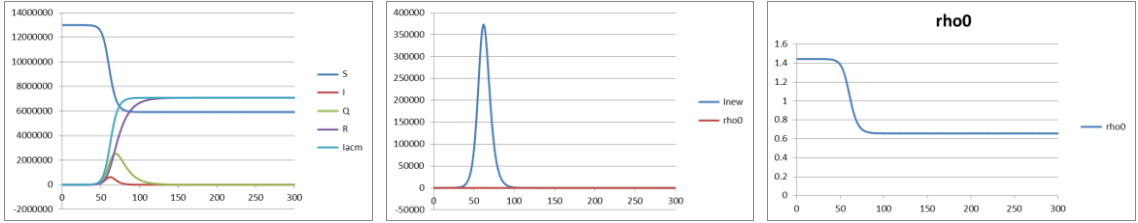

(a)  $S, I, Q, R, I_{acm}$      (b) New infectants      (c) Infection threshold

Fig. 4. When $Na$=13,000,000 (horizontal axis: time (day); vertical axis: number of people (person))

The effective infection opportunity population Na is significantly reduced to $N_a$=50,000 and the infection-force is set as $\beta$=0.655. The parameters used in the calculation are shown below:

Na=50,000 ; S[0]=49,999; I[0]=1; β=0.655; γ=0.08 ; q=0.45; q'=0.1

Checking the calculation results, the peak value of the newly infected person $I_{new}$ and the cumulative number of infected persons $I_{acm}$ are approaching to the observed values. But it still doesn't match. The period of infection is quite off.

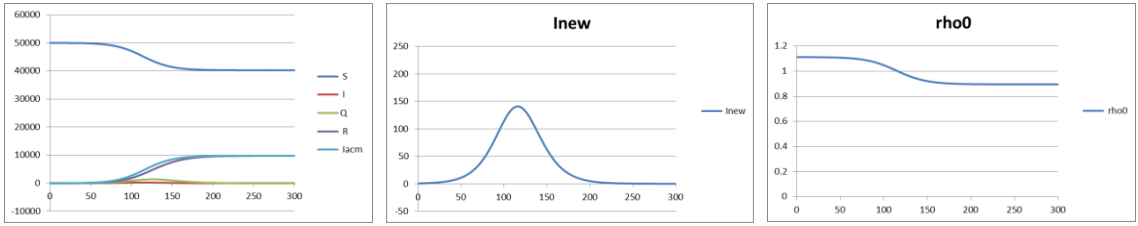

(a) $S, I, Q, R, I_{acm}$  (b) New infectants  (c) Infection threshold

Fig. 5. When $Na$=50,000 (horizontal axis: time (day); vertical axis: number of people (person))

The effective infection opportunity population is further reduced to $N_a$=25,000, and the infection-force is set as $\beta$=0.7. The parameters used in the calculation are shown below:

Na=25,000 ; S[0]=24,999; I[0]=1; β=0. 7 ; γ=0.08 ; p=0.45; q=0.1

Checking the calculation results, the peak value of the newly infected person $I_{new}$ and the cumulative number of infected persons $I_{acm}$ have come closer to the observed value. However, it is not good enough. The infection period is still large.

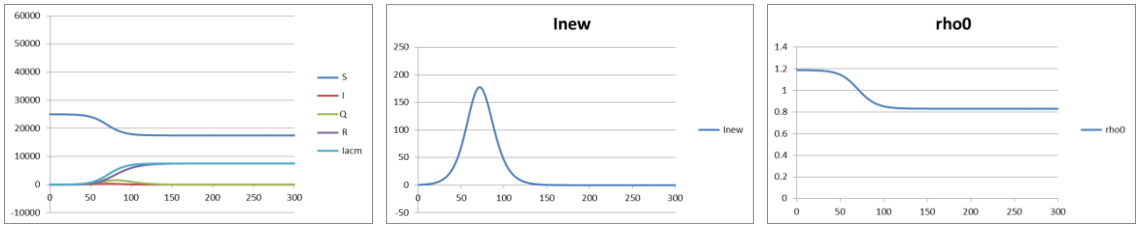

(a) $S, I, Q, R, I_{acm}$  (b) New infectants  (c) Infection threshold

Fig. 6. When $Na$=25,000 (horizontal axis: time (day); vertical axis: number of people (person))

The effective infection opportunity population is further reduced to $N_a$=10,000, and the infection-force is set as $\beta$=0.8. The parameters used in the calculation are shown below:

Na=10,000 ; S[0]=9,999; I[0]=1; β=0.8; γ=0.08 ; q=0.45; q'=0.1

Checking the calculation results, the target was fully achieved. The temporal aspects are also satisfactory.

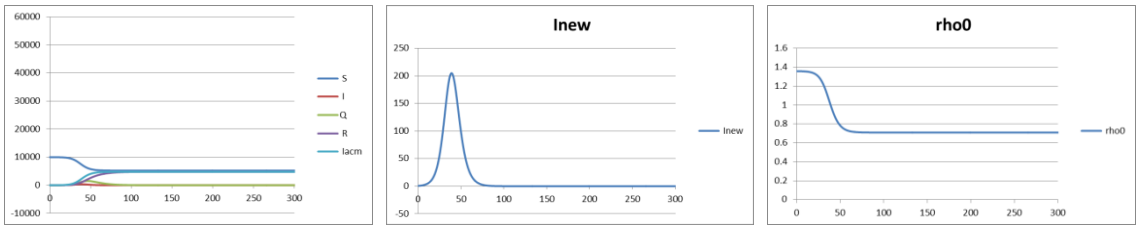

(a) $S, I, Q, R, I_{acm}$  (b) New infectants  (c) Infection threshold

Fig. 7. When $Na$=10,000 (horizontal axis: time (day); vertical axis: number of people (person))

## 5. Typical examples showing the dependency of infection on population

distribution

### 5.1. Calculation by the cellular automaton [4, 5, 6]

In the mean-field theory, the effects of the population distribution can't be reflected. However, in the case of the cellular automaton, the results differ if the population distribution is different. Since the infection occurs on the boundary of the colony of the infected people, the infection starts and terminates earlier in the distributed case than in the concentrated case as shown in Figs. 3 and 4 [6]. This should be considered a very important property of the population distribution. In the real infection phenomena, since the infection would spread from several distributed clusters of the infected people, the real infection spread quite differently to the SIR theory. Infection is also confined to a small number of people and most people remain unrelated.

The condition for the numerical calculation is shown below:

M=40, N=40, Whole population NT=1600 persons,

Period of calculation T=101days,

Recovery rate $\gamma$=0.08/day,

Initially uninfected people S(0)=1595persons,

Initially infected people I(0)=5person,

Initially recovered person R(0) =0person.

The numerical results by the 2D cellular automaton are shown below. For the simplicity of the algorithm, the recovered person is chosen probabilistically. In Figs. 8 and 9, ' S', 'X', and 'R' represent the uninfected, infected, and recovered people, respectively.

### 5.1.1. A case when initially infected persons are concentrated [6]

When the initially infected people exist concentratedly in the center of the population center, the infection spreads outwards towards the boundary. In Fig. 3, it is assumed that five infected persons appear at the center of the population space at t= 0.

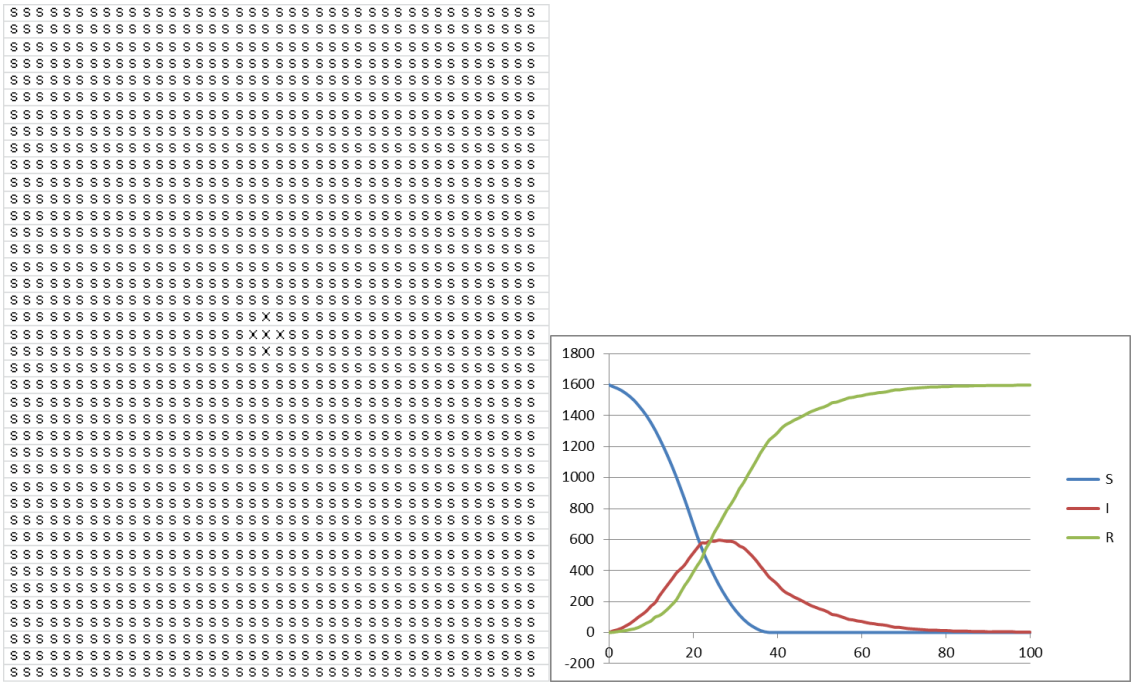

(a) The Initial state of cells (5 persons at center)   (b) Temporal change of *S*, *I*, and *R* (days vs. people)

Fig. 8. Numerical results 2 by a 2D cellular automaton A.

(Five initially infected persons are concentrated in the center).

### 5.1.2. A case when initially infected persons are dispersed [6]

When the initially infected persons are dispersed, the infection spreads multi-point and simultaneously. In Fig. 4, it is assumed that five infected persons are dispersed in space at t= 0.

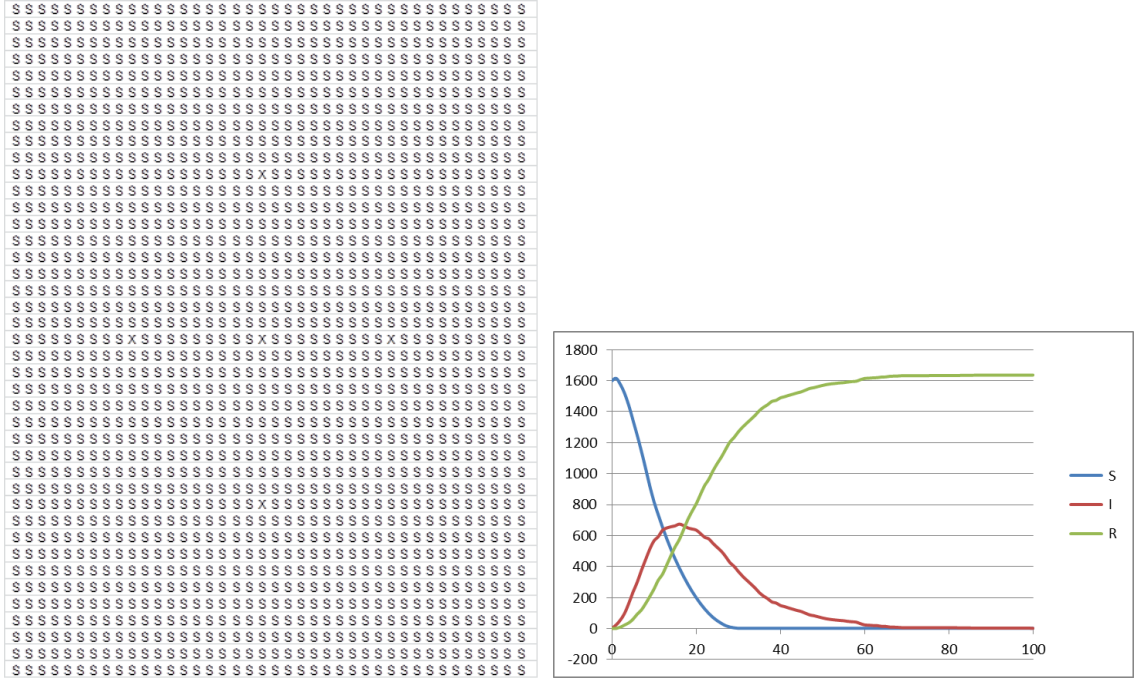

    The Initial state of cells (5 persons in space)    (b) Temporal change of S, I, and R (days vs. people)

Fig. 9. Numerical results 2 by a 2D cellular automaton B.

(Five initially infected persons are concentrated in the center).

Comparing the time evolutions of Figures 8 and 9, it can be seen that the spread of the initial infected person begins earlier and ends earlier than when the infected person is concentrated.

### 5.2 Comparison of numerical results by SIR and Extended SIR theories [7]

A comparison of the SIR and extended SIR is shown in Fig. 10. In this calculation, the conditions of both calculations are not equal. Namely, in the calculation using the extended SIR, the initially uninfected and infected persons are not distributed uniformly in the space, but one infected person appears at the center of the population space at t=0. In this case, both results are different. In the results by the extended SIR, the value of the peak is reduced and the position is delayed. The parameters for the SIR are set as

$$N_T=4410,\ T=100,\ dt=0.1,\ \beta=0.4,\ \gamma=0.08,\ S(0)=N_T-1,\ I(0)=1,\ R(0)=0. \tag{23}$$

and the parameters for the extended SIR are given by

$$N_T=4410,\ I_E=21,\ J_E=21,\ N[]= N_T/(I_E \times J_E)=10,\ T=100,\ dt=0.1,\ \beta=0.4,\ \gamma=0.08,\ t[][]=1.0,\ w[0][]=[1\ 1\ 1];$$
$$w[1][]=[1\ 0\ 1];\ w[2][]=[1\ 1\ 1]. \tag{24a}$$

The initial conditions are given below

$$S_{ij} = \begin{cases} \dfrac{N_T}{I_E J_E} - \dfrac{1}{I_E J_E} & \text{for } i=\dfrac{I_E}{2}+1,\ j=\dfrac{J_E}{2}+1 \\ \dfrac{N_T}{I_E J_E} & \text{otherwise} \end{cases},\quad I_{ij} = \begin{cases} \dfrac{1}{I_E J_E} & \text{for } i=\dfrac{I_E}{2}+1,\ j=\dfrac{J_E}{2}+1 \\ 0 & \text{otherwise} \end{cases},\quad R_{ij}=0.$$

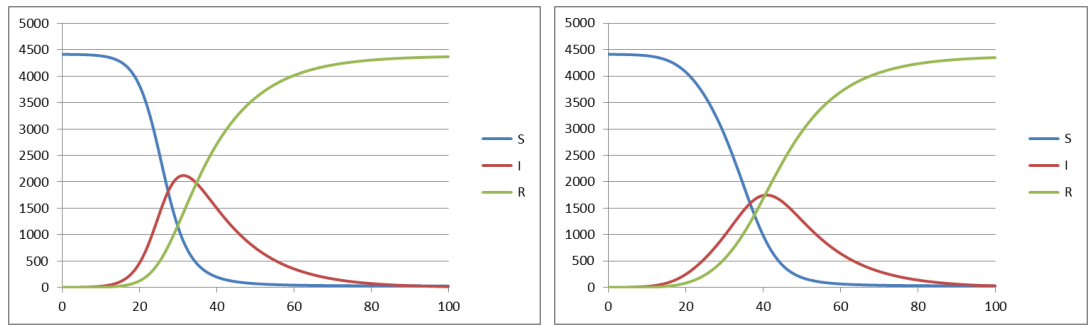

(a)　Results by SIR　　　　　　　　(b) Results by extended SIR

Fig. 10. Comparison 2 of numerical results by SIR and extended SIR theories

(Horizontal axis: Time,　Vertical axis: Number of people).

## 6. Conclusions

　　Since the new-type corona-virus takes a long time to develop, it is impossible to control it with any feedback-based measures that observe the current situation and take action. It can't be controlled unless it is a feed-forward proactive approach that predicts the future and dealing with it.

　　To do so, we must complete a mathematical theory that enables future prediction as soon as possible. Isn't this the only way to maintain economic activity while preventing corona's explosion during the epidemic?

## Acknowledgements

　　Mr. M. Suzuki, Dr. M. Hurutera, and Prof. H. Kajiwara are deeply acknowledged by the authors for their valuable suggestions and comments.

12